\documentclass[aps,prl,twocolumn,floatfix,showpacs,nofootinbib,superscriptaddress]{revtex4-1}

\usepackage{graphicx}
\usepackage{dcolumn}
\usepackage{amsmath}
\usepackage{longtable}
\usepackage{color}

\definecolor{purple}{rgb}{0.5,0,0.5}
\definecolor{blue}{rgb}{0.0,0,0.9}
\usepackage[colorlinks=true, pdfstartview=FitV, linkcolor=purple, citecolor= purple, urlcolor=blue]{hyperref}

\begin{document}
\title{Dressed-quark anomalous magnetic moments}

\author{Lei Chang}
\affiliation{Institute of Applied Physics and Computational
Mathematics, Beijing 100094, China}

\author{Yu-Xin Liu}
\affiliation{Department of Physics and State Key Laboratory of
Nuclear Physics and Technology, Peking University, Beijing 100871,
China}
\affiliation{Center of Theoretical Nuclear Physics, National
Laboratory of Heavy Ion Accelerator, Lanzhou 730000, China}

\author{Craig D.\ Roberts}
\affiliation{Department of Physics and State Key Laboratory of
Nuclear Physics and Technology, Peking University, Beijing 100871,
China}
\affiliation{Physics Division, Argonne National Laboratory, Argonne, Illinois 60439, USA}

\date{\today\ }

\begin{abstract}
Perturbation theory predicts that a massless fermion cannot possess a measurable magnetic moment.  We explain, however, that the nonperturbative phenomenon of dynamical chiral symmetry breaking generates a momentum-dependent anomalous chromomagnetic moment for dressed light-quarks, which is large at infrared momenta; and demonstrate that consequently these same quarks also possess an anomalous electromagnetic moment with similar magnitude and opposite sign.
\end{abstract}

\pacs{
12.38.Aw,   
12.38.Lg,   
11.30.Rd,   
24.85.+p    
}

\maketitle

In Dirac's relativistic quantum mechanics, a fermion with charge $q$ and mass $m$, interacting with an electromagnetic field, has a magnetic moment\footnote{We use \emph{natural units}, $\hbar =1 =c $, and a Euclidean metric: $\{\gamma_\mu,\gamma_\nu\} = 2\delta_{\mu\nu}$; $\gamma_\mu^\dagger = \gamma_\mu$; $\sigma_{\mu\nu}=(i/2)[\gamma_\mu,\gamma_\nu]$; $a \cdot b = \sum_{i=1}^4 a_i b_i$; and $P_\mu$ timelike $\Rightarrow$ $P^2<0$.} $\mu= q/[2 m]$.  This prediction held true for the electron until improvements in experimental techniques enabled the discovery of a small deviation \cite{Foley:1948zz}, with the moment increased by a multiplicative factor: $1.00119\pm 0.00005$.  This correction was explained by the first systematic computation using renormalized quantum electrodynamics (QED) \cite{Schwinger:1948iu}:
\begin{equation}
\label{anommme}
\frac{q}{2m} \to \left(1 + \frac{\alpha}{2\pi}\right) \frac{q}{2m}\,,
\end{equation}
where $\alpha$ is QED's fine structure constant.  The agreement with experiment established quantum electrodynamics as a valid tool.  The correction defines the electron's \emph{anomalous magnetic moment}, which is now known with extraordinary precision and agrees with theory at O$(\alpha^5)$ \cite{Mohr:2008fa}.

The fermion-photon coupling in QED is described by:
\begin{equation}
\label{QEDinteraction}
\int d^4\! x\, i q \,\bar\psi(x) \gamma_\mu \psi(x)\,A_\mu(x)\,,
\end{equation}
This interaction generates the following electromagnetic current for an on-shell Dirac fermion ($k=p_f -p_i$),
%
\begin{equation}
i q \, \bar u(p_f) \left[ \gamma_\mu F_1(k^2)+ \frac{1}{2 m} \sigma_{\mu\nu} k_\nu F_2(k^2)\right] u(p_i)\,,
\end{equation}
where: $F_1(k^2)$, $F_2(k^2)$ are form factors; and $u(p)$, $\bar u(p)$ are spinors, 
the free particle forms of which satisfy
\begin{equation}
\label{Diracm}
\bar u(p_f) ( i\gamma\cdot p_f + m ) = 0 \,,\;
( i\gamma\cdot p_i + m ) u(p_i) = 0\,.
\end{equation}

A Gordon-identity can be obtained from these ``Dirac'' equations; viz., with $2 \ell=p_f + p_i$,
\begin{equation}
\label{Gordon}
2 m \, \bar u(p_f) i \gamma_\mu u(p_i) = \bar u(p_f)\left[ 2 \ell_\mu + i \sigma_{\mu\nu} k_\nu \right]u(p_i)\,.
\end{equation}
It follows that a point-particle in the absence of radiative corrections has $F_1 \equiv 1$ and $F_2 \equiv 0$, and hence Dirac's value for the magnetic moment.  The anomalous magnetic moment in Eq.\,(\ref{anommme}) corresponds to $F_2(0) = \alpha/2\pi$.

An anomalous contribution to the moment can therefore be associated with an additional interaction term:
\begin{equation}
\label{anominteraction}
\int d^4\! x\, \mbox{\small $\frac{1}{2}$} q \, \bar \psi(x) \sigma_{\mu\nu} \psi(x) F_{\mu\nu}(x)\,,
\end{equation}
where $F_{\mu\nu}(x)$ is the gauge-boson field strength tensor.  This term is invariant under local $U(1)$ gauge transformations but is not generated by minimal substitution in the action for a free Dirac field.

Consider the effect of the global chiral transformation $\psi(x) \to \exp(i \theta\gamma_5) \psi(x)$.
The term in Eq.\,(\ref{QEDinteraction}) is invariant.
However, the interaction of Eq.\,(\ref{anominteraction}) is not.  These observations facilitate the understanding of a general result: $F_2\equiv 0$ for a massless fermion in a quantum field theory with chiral symmetry realized in the Wigner mode; i.e., when the symmetry is not dynamically broken.
A firmer conclusion can be drawn.  For $m=0$ it follows from Eq.\,(\ref{Gordon}) that Eq.\,(\ref{QEDinteraction}) does not mix with the helicity-flipping interaction of Eq.\,(\ref{anominteraction}) and hence a massless fermion does not possess a measurable magnetic moment.

A reconsideration of Ref.\,\cite{Schwinger:1948iu} reveals no manifest conflict with these facts.  The perturbative expression for $F_2(0)$ contains a multiplicative numerator factor of $m$ and the usual analysis of the denominator involves steps that are only valid for $m\neq 0$.  Fundamentally, there is no conundrum because QED is not an asymptotically free theory and hence does not possess a well-defined chiral limit.

On the other hand, in quantum chromodynamics (QCD)
the chiral limit is rigorously defined nonperturbatively \cite{Maris:1997hd,Maris:1997tm}.  This non-Abelian quantum field theory describes quarks interacting via the exchange of gluons, which themselves self-interact.  Apart from the inclusion of a matrix to represent the color degree of freedom, the quark-gluon interaction is described by Eq.\,(\ref{QEDinteraction}).  The analogue of Schwinger's one-loop calculation can be carried out to find an anomalous \emph{chromo}-magnetic moment for the quark.  There are two diagrams in this case: one similar in form to that in QED; and another owing to the gluon self-interaction.  One reads from Ref.\,\cite{Davydychev:2000rt} that the perturbative result vanishes in the chiral limit.
However, nonperturbative studies of QCD's gap equation \cite{Bhagwat:2003vw} and numerical simulations of the lattice-regularized theory \cite{Bowman:2005vx} have revealed that chiral symmetry is dynamically broken in QCD.
Does this affect the chromomagnetic moment?

Dynamical chiral symmetry breaking (DCSB) is a remarkably effective mass generating mechanism, which can be explained via the dressed-quark propagator
\begin{equation}
\label{Spzeta}
S(p;\zeta) = 1/[i\gamma\cdot p A(p^2;\zeta) + B(p^2;\zeta)]\,,
\end{equation}
where $\zeta$ is the renormalization mass-scale and the dressed-quark mass function $M(p^2)=B(p^2;\zeta)/A(p^2;\zeta)$ is renormalization point invariant.  In the chiral limit,
$M(p^2)$ is identically zero at any finite order in perturbation theory.  However, DCSB generates mass \emph{from nothing}.  Thus, in chiral-QCD, $M(p^2)$ is strongly momentum-dependent and $M(p^2) \approx 0.5\,$GeV.  DCSB is the origin of constituent-quark masses 
and intimately connected with confinement in QCD \cite{Chang:2010jq}.

QCD dynamics, and DCSB in particular, also have a material effect on the quark-gluon vertex:
\begin{equation}
\label{dqgv}
\Gamma_\mu^a(p_f,p_i;k) = \frac{\lambda^a}{2} \, \Gamma_\mu(p_f,p_i;k)\,,
\end{equation}
where $\{\lambda^a|a=1,\ldots,8\}$ are the color Gell-Mann matrices.  $\Gamma_\mu(p_f,p_i)$ can be expressed via twelve independent Dirac-matrix-valued tensors, each multiplied by a scalar function.  It has long been known via Dyson-Schwinger equation (DSE) studies \cite{Roberts:1994dr} that at least three of the tensors are materially modified from their perturbative forms in strongly interacting theories; namely, $\lambda_{1,2,3}$ in
\begin{eqnarray}
\nonumber
\lefteqn{i \Gamma_\mu(p_f,p_i;k) = \lambda_1(p_f,p_i;k) i \gamma_\mu }\\
&& +
2 \ell_\mu \left[i\gamma\cdot \ell \,\lambda_2(p_f,p_i;k)
+
\lambda_3(p_f,p_i;k)\right] + [\ldots].
\end{eqnarray}
These terms constitute the so-called longitudinal vertex and are constrained by the Slavnov-Taylor identity, a non-Abelian form of the Ward-Takahashi identity.

Contemporary simulations of lattice-regularized QCD \cite{Skullerud:2003qu} and DSE studies \cite{Bhagwat:2004kj} agree that
\begin{equation}
\label{alpha3B}
\lambda_3(p,p;0) \approx \frac{d}{dp^2} B(p^2,\zeta)
\end{equation}
and also on the form of $\lambda_1(p,p;0)$, which is functionally similar to $A(p^2,\zeta)$.  However, owing to non-orthogonality of the tensors accompanying $\lambda_1$ and $\lambda_2$, it is difficult to obtain a lattice signal for $\lambda_2$.  We therefore consider the DSE prediction in Ref.\,\cite{Bhagwat:2004kj} more reliable.

Perturbative massless-QCD conserves helicity so the quark-gluon vertex cannot perturbatively have a term with the helicity-flipping characteristics of $\lambda_3$.  Equation~(\ref{alpha3B}) is thus remarkable, showing that the dressed-quark-gluon vertex contains at least one chirally-asymmetric component whose origin and size owe solely to DCSB.  A recent advance in understanding the Bethe-Salpeter equation (BSE) has enabled practitioners to establish that $\lambda_3$ has a big impact on the hadron spectrum \cite{Chang:2009zb}; e.g., it generates a very strong spin-orbit interaction.

We will take this reasoning further.  As explained above, massless fermions in gauge field theories cannot possess an anomalous chromo/electro-magnetic moment because the term that describes it couples left- and right-handed fermions.  However, if chiral symmetry is strongly broken dynamically, then the fermions should also posses large anomalous magnetic moments.  Such an effect is expressed in the dressed-quark-gluon vertex via a term
\begin{equation}
\label{qcdanom1}
\Gamma_\mu^{\rm acm_5} (p_f,p_i;k) = \sigma_{\mu\nu} k_\nu \, \tau_5(p_f,p_i,k)\,.
\end{equation}

That QCD generates a strongly momentum-dependent chromomagnetic form factor in the quark-gluon vertex, $\tau_5$, with a large DCSB-component, is confirmed in Ref.\,\cite{Skullerud:2003qu}.  Only a particular kinematic arrangement was readily accessible in that lattice simulation but this is enough to learn that, at the current-quark mass considered: $\tau_5$ is roughly two orders-of-magnitude larger than the perturbative form; and
\begin{equation}
\label{boundtau5}
\forall p^2>0: \; |\tau_5(p,-p;2 p)| \gtrsim |\lambda_3(p,p;0)|\,.
\end{equation}
The magnitude of the lattice result is consistent with instanton-liquid model estimates \cite{Kochelev:1996pv,Diakonov:2002fq}.

This large chromomagnetic moment is likely to have a broad impact on the properties of light-quark systems \cite{Diakonov:2002fq,Ebert:2005es}.  In particular, it can probably explain the longstanding puzzle of the mass splitting between the $a_1$- and $\rho$-mesons in the hadron spectrum \cite{Chang:2010jq}.  Herein, however, we will elucidate another novel effect; viz., the manner in which the quark's chromomagnetic moment generates a quark anomalous \emph{electro}magnetic moment.  The method of Ref.\,\cite{Chang:2009zb} makes this possible for the first time.

Following Ref.\,\cite{Chang:2009zb}, one need only specify the gap equation's kernel because the quark-photon vertex BSE  is completely defined therefrom.  The gap equation is
\begin{eqnarray}
\nonumber S(p)^{-1} &=& Z_2 \,(i\gamma\cdot p + m^{\rm bm}) + Z_1 \int^\Lambda_k\!\! g^2 D_{\mu\nu}(p-k) \\
&& \times \frac{\lambda^a}{2}\gamma_\mu S_f(q) \frac{\lambda^a}{2}\Gamma_\nu(k,p) ,
\label{gendse}
\end{eqnarray}
where: $D_{\mu\nu}$ is the gluon propagator; $\Gamma_\nu$ is the quark-gluon vertex, Eq.\,(\ref{dqgv}); $\int^\Lambda_k:=\int^\Lambda d^4k/(2\pi)^4$ is a Poincar\'e invariant regularization of the integral, with $\Lambda$ the regularization mass-scale; $m^{\rm bm}(\Lambda)$ is the Lagrangian bare mass; and $Z_{1,2}(\zeta^2,\Lambda^2)$ are renormalization constants.  

The kernel can be rendered tractable by writing \cite{Maris:1997tm}
\begin{equation}
Z_1 g^2 D_{\rho \sigma}(t) \Gamma_\sigma(q,q+t)
= {\cal G}(t^2) \, D_{\rho\sigma}^{\rm free}(t) \tilde\Gamma_\sigma(q,q+t)\,, \label{KernelAnsatz}
\end{equation}
wherein $D_{\rho \sigma}^{\rm free}$ is the Landau-gauge free-gauge-boson propagator, ${\cal G}$ is an interaction model and $\tilde\Gamma_\sigma$ is an \emph{Ansatz} for the quark-gluon vertex.  For the interaction, we use
\begin{equation}
\label{IRGs}
{\cal G}(\ell^2) = \frac{4\pi^2}{\omega^6}\, D\, \ell^4\, {\rm e}^{-\ell^2/\omega^2},
\end{equation}
a simplified form of that in Ref.\,\cite{Maris:1997tm}.  This enables us to avoid renormalization, which is straightforward but not germane to an analysis of vertex contributions that are power-law suppressed in the ultraviolet.

In order to explain the vertex \emph{Ansatz} to be used, we return to perturbation theory.  One can determine from Ref.\,\cite{Davydychev:2000rt} that at leading-order in the coupling, $\alpha_s$, the three-gluon vertex does not contribute to the QCD analogue of Eq.\,(\ref{anommme}) and the Abelian-like diagram produces the finite and negative correction $(-\alpha_s/[12 \pi])$.
The complete cancelation of ultraviolet divergences occurs only because of the dynamical generation of another term in the transverse part of the quark-gluon vertex; namely,
\begin{equation}
\Gamma_\mu^{\rm acm_4}(p_f,p_i) = [ \ell_\mu^{\rm T} \gamma\cdot  k + i \gamma_\mu^{\rm T} \sigma_{\nu\rho}\ell_\nu k_\rho] \tau_4(p_f,p_i)\,,
\end{equation}
with $T_{\mu\nu} = \delta_{\mu\nu} - k_\mu k_\nu/k^2$, $a_\mu^{\rm T} := T_{\mu\nu}a_\nu$.

Cognisant of this, we use a simple \emph{Ansatz} to express the dynamical generation of an anomalous chromomagnetic moment via the dressed-quark gluon vertex; viz.,
\begin{eqnarray}
\label{ourvtx}
\tilde\Gamma_\mu(p_f,p_i)  & = & \Gamma_\mu^{\rm BC}(p_f,p_i) +
\Gamma_\mu^{\rm acm}(p_f,p_i)\,,\\
\nonumber
i\Gamma_\mu^{\rm BC}(p_f,p_i)  & = &
i\Sigma_A(p_f^2,p_i^2)\,\gamma_\mu + 2 \ell_\mu \left[ i\gamma\cdot \ell \,\Delta_A(p_f^2,p_i^2)  \right. \\
&&  \left. + \Delta_B(p_f^2,p_i^2)\right] ,
\label{bcvtx}
\end{eqnarray}
where \cite{Ball:1980ay} $\Sigma_{\phi}(p_f^2,p_i^2) = [\phi(p_f^2)+\phi(p_i^2)]/2$, $\Delta_{\phi}(p_f^2,p_i^2) = [\phi(p_f^2)-\phi(p_i^2)]/[p_f^2-p_i^2]$, and
\begin{equation}
\Gamma_\mu^{\rm acm}(p_f,p_i) = \Gamma_\mu^{\rm acm_4}(p_f,p_i) + \Gamma_\mu^{\rm acm_5}(p_f,p_i)\,,
\end{equation}
with $\tau_5(p_f,p_i) =  \eta\Delta_B(p_f^2,p_i^2)$, as discussed above, and
\begin{equation}
\tau_4(p_f,p_i) = {\cal F}(z) \bigg[  \frac{1-2\eta}{M_E}\Delta_B(p_f^2,p_i^2) - \Delta_A(p_f^2,p_i^2) \bigg]. \label{tau4}
\end{equation}
The damping factor ${\cal F}(z)=(1- \exp(-z))/z$, $z=(p_i^2 + p_f^2- 2 M_E^2)/\Lambda_{\cal F}^2$, $\Lambda_{\cal F}=1\,$GeV, simplifies numerical analysis; and $M_E=\{ s| s>0, s = M^2(s)\}$ is the Euclidean constituent-quark mass.
A realistic description of the light-quark meson spectrum is obtained with $\omega=0.5\,$GeV, $D=(0.72\,{\rm GeV})^2$, $m=5\,$MeV, $\eta=-7/4$.

A confined quark does not possess a mass-shell \cite{Roberts:1994dr,Roberts:2007ji}.  Hence, one cannot unambiguously assign a single value to its anomalous magnetic moment.  One can nonetheless compute a magnetic moment distribution.  At each value of $p^2$, we define spinors to satisfy Eqs.\,(\ref{Diracm}) with $m\to M(p^2)=:\varsigma$, and use
\begin{eqnarray}
\nonumber
\lefteqn{\bar u(p_f;\varsigma) \, \Gamma_\mu( p_f,p_i;k)\,  u(p_i;\varsigma)}\\
& = &
\bar u(p_f) [ F_1(k^2) \gamma_\mu + \frac{1}{2 \varsigma} \,\sigma_{\mu \nu} k_\nu F_2(k^2)] u(p_i).
\label{GenSpinors}
\end{eqnarray}
Now, from Eqs.\,(\ref{ourvtx}) -- (\ref{tau4}), one finds
\begin{equation}
\label{kappaacm}
\kappa^{\rm acm}(\varsigma) = \frac{ - 2 \varsigma \, \eta \delta_B^{\varsigma}}
    {\sigma_A^{\varsigma} - 2 \varsigma^2 \delta_A^{\varsigma}+ 2 \varsigma \delta_B^{\varsigma} }\,,
\end{equation}
where $\sigma_A^{\varsigma} = \Sigma_A(\varsigma,\varsigma)$, $\delta_A^{\varsigma} = \Delta_A(\varsigma,\varsigma)$, etc.  The numerator's simplicity owes to a premeditated cancelation between $\tau_4$ and $\tau_5$, which replicates the one at leading-order in perturbation theory.
Where a comparison of terms is possible, our vertex \emph{Ansatz} is semi-quantitatively in agreement with Refs.\,\cite{Skullerud:2003qu,Bhagwat:2004kj}.  However, the presence and understanding of the role of $\Gamma_\mu^{\rm acm_4}$ is novel.
%
(NB.\ It is apparent from Eq.\,(\ref{kappaacm}) that $\kappa^{\rm acm} \propto m^2$ in the absence of DCSB, so that $\kappa^{\rm acm}/[2m]\to 0$ in the chiral limit.)

We can now write the BSE for the quark-photon vertex following the method of Ref.\,\cite{Chang:2009zb}.  This is nontrivial but details will be reported elsewhere.  Since the method guarantees preservation of the Ward-Takahashi identities, the general form of the solution is
\begin{eqnarray}
\Gamma_\mu^\gamma(p_f,p_i) & = & \Gamma_\mu^{\rm BC}(p_f,p_i) + \Gamma_\mu^{\rm T}(p_f,p_i)\,,\\
\nonumber
\Gamma_\mu^{\rm T}(p_f,p_i) & = &
\gamma_\mu^{\rm T} \hat F_1
+ \sigma_{\mu\nu} k_\nu \hat F_2
+ T_{\mu\rho} \sigma_{\rho\nu} \ell_\nu \,\ell\cdot k\, \hat F_3\\
\nonumber & &
+ [ \ell_\mu^{\rm T} \gamma\cdot  k + i \gamma_\mu^{\rm T} \sigma_{\nu\rho}\ell_\nu k_\rho] \hat F_4 - i \ell_\mu^{\rm T} \hat F_5\\
\nonumber & & +\, \ell_\mu^{\rm T} \gamma\cdot k \, \ell \cdot  k\, \hat F_6 - \ell_\mu^{\rm T} \gamma\cdot \ell \, \hat F_7  \\
%
& & + \ell_\mu^T \sigma_{\nu\rho} \ell_\nu k_\rho \hat F_8
\end{eqnarray}
and $\{F_i|i=1,\ldots,8\}$ are scalar functions. 
The Ward-Takahashi identity is plainly satisfied; viz., $k_\mu i \Gamma_\mu(p_f,p_i)
= 1/S(p_f) - 1/S(p_i)$.

\begin{figure}[t]
\vspace*{-5ex}

\centerline{\includegraphics[clip,width=0.43\textwidth]{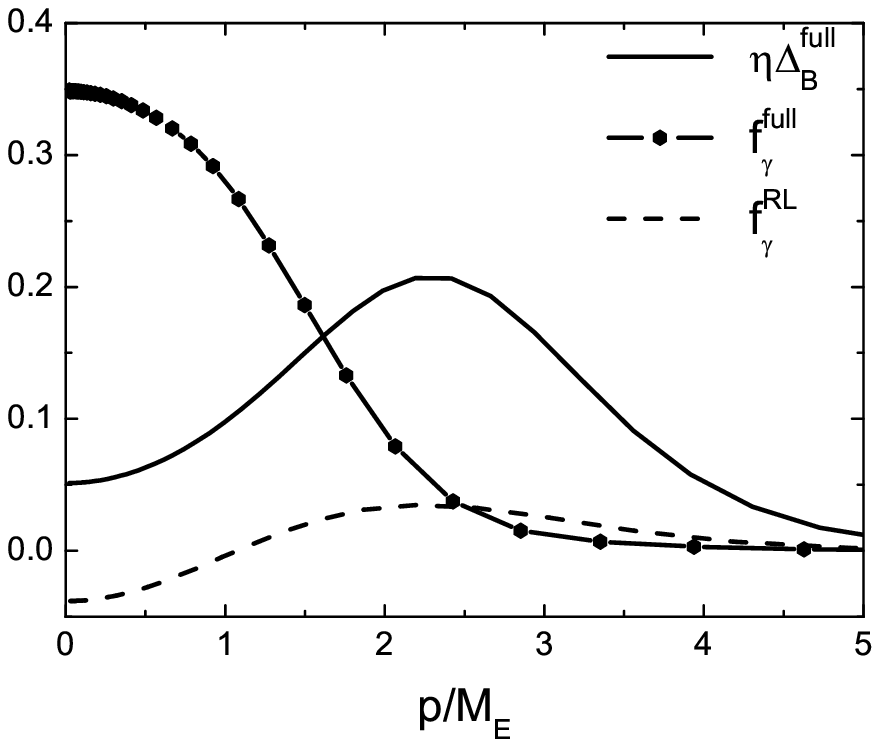}}
\vspace*{-6ex}

\centerline{\includegraphics[clip,width=0.45\textwidth]{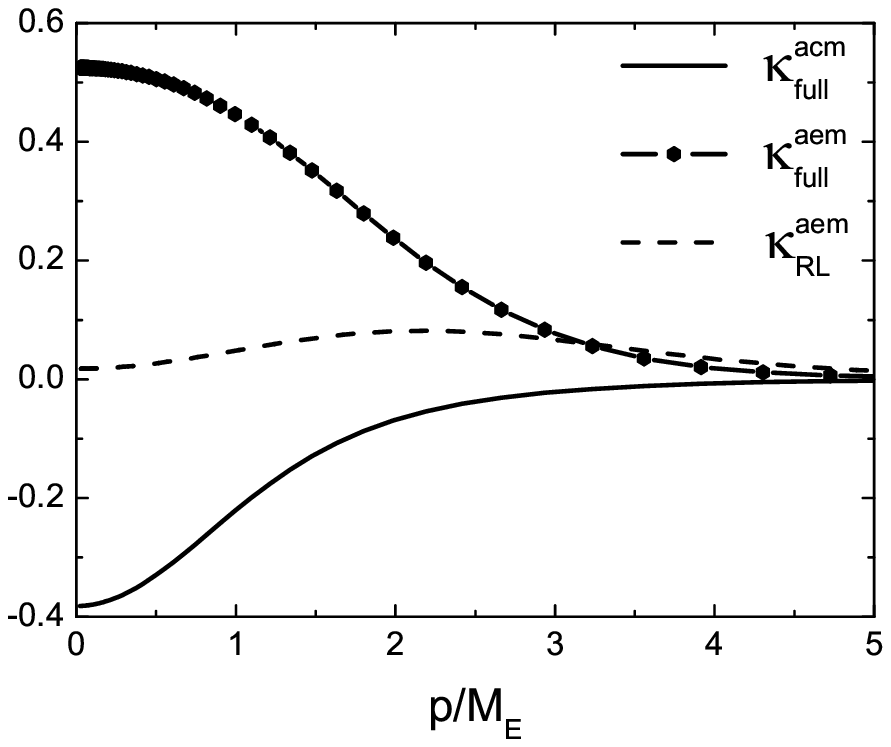}\rule{1em}{0em}}
\vspace*{-5ex}

\caption{\label{figACM}
\emph{Upper panel} -- $f_\gamma$ (GeV$^{-1}$) in Eq.\,(\protect\ref{fgamma})
cf.\ $\eta\Delta_B(p^2,p^2)$, both computed using Eqs.\,(\protect\ref{IRGs}), (\protect\ref{ourvtx}).
\emph{Lower panel} -- Anomalous chromo- and electro-magnetic moment distributions for a dressed-quark, computed using Eq.\,(\protect\ref{kappavalue}).
The dashed-curve in both panels is the rainbow-ladder (RL) truncation result.
}

\end{figure}

We have solved for the vertex and computed the quark's anomalous electromagnetic moment form factor
\begin{equation}
f_\gamma(p) :=  \lim_{p_f\to p}\frac{-1}{12\,k^2}{\rm tr}\, \sigma_{\mu\nu} k_\mu \Gamma_\nu^\gamma(p_f,p)
=  \hat F_2+ \frac{1}{3}p^2 \hat F_8\,. \label{fgamma}
\end{equation}
The result is sizable, Fig.\,\ref{figACM}.
We reiterate that $f_\gamma$ is completely nonperturbative: in the chiral limit, at any finite order in perturbation theory, $f_\gamma\equiv 0$, both in our model and in QCD.
For contrast we also plot the result obtained in the rainbow-ladder truncation of QCD's DSEs. As the leading-order in a systematic but stepwise symmetry-preserving scheme \cite{Bender:1996bb}, this truncation only partially expresses DCSB: it is exhibited by the dressed-quark propagator but not present in the quark-gluon vertex.  In this case $f_\gamma$ is nonzero but small.  These are artefacts of the truncation that will not be remedied at any finite order of the procedure in Ref.\,\cite{Bender:1996bb} or a kindred scheme.

Employing Eq.\,(\ref{GenSpinors}),
one can write an expression for the quark's anomalous electromagnetic moment distribution
\begin{equation}
\label{kappavalue}
\kappa(\varsigma)=\frac{2 \varsigma \hat F_{2} + 2 \varsigma^2 \hat F_4  +\Lambda_{\kappa}(\varsigma)}
{\sigma_{A}^{\varsigma} + \hat F_{1}-\Lambda_{\kappa}(\varsigma)}\,,
\end{equation}
where: $\Lambda_{\kappa}(\varsigma)= 2\varsigma^{2}\delta_{A}^\varsigma-2 \varsigma \delta_{B}^\varsigma -\varsigma \hat F_5 - \varsigma^2 \hat F_7$; and the $\hat F_i$ are evaluated at $p_f^2=p_i^2=M(p_f^2)^2=:\varsigma^2$, $k^2=0$.  Plainly, $\kappa(\varsigma)\equiv 0$ in the chiral limit when chiral symmetry is not dynamically broken.  Moreover, as a consquence of asymptotic freedom, $\kappa(\varsigma) \to 0$ rapidly with increasing momentum.
Our computed distribution is depicted in Fig.\,\ref{figACM}.  It yields Euclidean mass-shell values: $M_{\rm full}^E = 0.44\,$GeV, $\kappa_{\rm full}^{\rm acm}= -0.22\,$, $\kappa_{\rm full}^{\rm aem}= 0.45\,$ cf.\ $M_{\rm RL}^E = 0.35\,$GeV, $\kappa_{\rm RL}^{\rm acm}= 0\,$, $\kappa_{\rm RL}^{\rm aem}= 0.048$.

We explained how dynamical chiral symmetry breaking produces a dressed light-quark with a momentum-dependent anomalous chromomagnetic moment, which is large at infrared momenta and whose existence is likely to have many observable consequences.  Significant amongst them is the generation of an anomalous electromagnetic moment for the dressed light-quark with commensurate size but opposite sign.  The infrared scale of both moments is determined by the Euclidean constituent-quark mass.  This is two orders-of-magnitude greater than the physical light-quark current-mass, which sets the scale of the perturbative result for both these quantities.  For the hadron physics practitioner, there are two additional notable features; namely, the rainbow-ladder truncation, and low-order stepwise improvements thereof, underestimate these effects by an order of magnitude; and both the $\tau_4$ and $\tau_5$ terms in the dressed-quark-gluon vertex are indispensable for a realistic description of hadron phenomena.  Whilst we used a simple interaction to illustrate these outcomes, they are robust.

Our results should stimulate and provide the basic input for a reanalysis of the hadron spectrum and hadron elastic and transition electromagnetic form factors with these novel effects taken into account.  Furthermore, given the magnitude of the muon ``$g_\mu-2$ anomaly'' and its assumed importance as an harbinger of physics beyond the Standard Model \cite{Bennett:2006fi}, it might also be worthwhile to make a quantitative estimate of the contribution to $g_\mu-2$ from the quark's DCSB-induced anomalous moments.  These contributions appear in the hadronic component of the photon polarization tensor.

%
We are grateful for comments from A.~K{\i}z{\i}lers\"u and P.\,C.~Tandy.
This work was funded by:
the National Natural Science Foundation of China, contract
nos.~10425521, 10705002, 10935001;  
%
%
the Major State Basic Research Development Program, contract no.~G2007CB815000;
and the United States Department of Energy, Office of Nuclear Physics, contract no.~DE-AC02-06CH11357.
%



\begin{thebibliography}{00}


\bibitem{Foley:1948zz}
  H.~M.~Foley and P.~Kusch,
  Phys.\ Rev.\  {\bf 73}, 412 (1948).

\bibitem{Schwinger:1948iu}
  J.~S.~Schwinger,
  Phys.\ Rev.\  {\bf 73}, 416 (1948).

\bibitem{Mohr:2008fa}
  P.~J.~Mohr, B.~N.~Taylor and D.~B.~Newell,
  Rev.\ Mod.\ Phys.\  {\bf 80}, 633 (2008).

\bibitem{Maris:1997hd}
  P.~Maris, C.~D.~Roberts and P.~C.~Tandy,
  Phys.\ Lett.\  B {\bf 420}, 267 (1998).

\bibitem{Maris:1997tm}
  P.~Maris and C.~D.~Roberts,
  Phys.\ Rev.\  C {\bf 56}, 3369 (1997).

\bibitem{Davydychev:2000rt}
  A.~I.~Davydychev, P.~Osland and L.~Saks,
  Phys.\ Rev.\  D {\bf 63}, 014022 (2001).

\bibitem{Bhagwat:2003vw}
  M.~S.~Bhagwat, M.~A.~Pichowsky, C.~D.~Roberts and P.~C.~Tandy,
  Phys.\ Rev.\  C {\bf 68}, 015203 (2003);
  M.~S.~Bhagwat and P.~C.~Tandy,
  AIP Conf.\ Proc.\  {\bf 842}, 225 (2006).

\bibitem{Bowman:2005vx}
  P.~O.~Bowman \emph{et al}., 
  Phys.\ Rev.\  D {\bf 71}, 054507 (2005).

\bibitem{Roberts:1994dr}
  C.~D.~Roberts and A.~G.~Williams,
  Prog.\ Part.\ Nucl.\ Phys.\  {\bf 33}, 477 (1994).

\bibitem{Chang:2010jq}
  L.~Chang and C.~D.~Roberts,
  ``Hadron Physics: The Essence of Matter,''
  arXiv:1003.5006 [nucl-th].

\bibitem{Skullerud:2003qu}
  J.~I.~Skullerud \emph{et al}., 
  JHEP {\bf 0304}, 047 (2003).

\bibitem{Bhagwat:2004kj}
  M.~S.~Bhagwat and P.~C.~Tandy,
  Phys.\ Rev.\  D {\bf 70}, 094039 (2004).

\bibitem{Chang:2009zb}
  L.~Chang and C.~D.~Roberts,
  Phys.\ Rev.\ Lett.\  {\bf 103}, 081601 (2009).

\bibitem{Kochelev:1996pv}
  N.~I.~Kochelev,
  Phys.\ Lett.\  B {\bf 426}, 149 (1998).

\bibitem{Diakonov:2002fq}
  D.~Diakonov,
  Prog.\ Part.\ Nucl.\ Phys.\  {\bf 51}, 173 (2003).

\bibitem{Ebert:2005es}
  D.~Ebert, R.~N.~Faustov and V.~O.~Galkin,
  Eur.\ Phys.\ J.\  C {\bf 47}, 745 (2006).

\bibitem{Ball:1980ay}
  J.~S.~Ball and T.~W.~Chiu,
  Phys.\ Rev.\  D {\bf 22}, 2542 (1980).

\bibitem{Roberts:2007ji}
  C.~D.~Roberts,
  Prog.\ Part.\ Nucl.\ Phys.\  {\bf 61}, 50 (2008).

\bibitem{Bender:1996bb}
  A.~Bender, C.~D.~Roberts and L.~Von Smekal,
  Phys.\ Lett.\  B {\bf 380}, 7 (1996).

\bibitem{Bennett:2006fi}
  G.~W.~Bennett {\it et al.},  
  Phys.\ Rev.\  D {\bf 73}, 072003 (2006).


\end{thebibliography}
\end{document}